\documentclass[referee]{aa} % for a referee version
\usepackage{graphicx}
\usepackage{txfonts}
\begin{document}
   \title{ Modeling Star counts in the Monoceros stream and the Galactic anti-centre.
   }

   \subtitle{ }

   \author{P. L. Hammersley
          \inst{1,2,3}
          \and
         M. L\'opez-Corredoira \inst{1,2}
          }

   \institute{       Instituto de Astrof\'\i sica de Canarias, E-38200  La Laguna,  Tenerife, Spain.\\
              \email{}
         \and
             Departamento de Astrof\'\i sica, Universidad de La Laguna,  La Laguna,  Tenerife, Spain\\
	     \email{peter.hammersley@iac.es, martinlc@iac.es }
	   \and 
	      European Southern Observatory, Karl-Schwarzschild-Strasse 2, Garching 85748, Germany 
             }
 
   \date{}

% \abstract{}{}{}{}{} 
% 5 {} token are mandatory
 
  \abstract
    { There is a continued debate as to the form of the outer disc of the
    Milky Way galaxy, which has important implications for its formation. Stars are known to exist at a galacto-centric distance
    of at least 20~kpc. However, there is much debate as to whether  these stars can be explained as being part of the disc or
    whether another extra galactic structure, the so called   Monoceros ring/stream, is  required}  
  {To examine the outer disc of the Galaxy toward the anti-centre 
  to determine whether the star counts
   can be explained by the thin and thick discs alone.}
  {Using Sloan star counts and extracting the late F and early G dwarfs it is
   possible to directly determine  the density of stars out to a galacto-centric distance of
   about 25~kpc. These are then compared with a simple flared disc model.}
  { A flared  disc model is shown to reproduce 
   the counts along the line of sights examined, if the thick disc does not have a sharp cut off.  
   The flare starts at a Galacto-centric radius of 16~kpc and has a scale length of
   4.5 $\pm$1.5 kpc}
  { Whilst the interpretation of the counts in terms of a ring/stream cannot be definitely discounted, 
it does not appear to be necessary, at least along the lines of sight examined towards the anti centre.}
   {}


   \maketitle
%
%__________________________________________________________outer_new.tex______

\section{Introduction}

 The position of the Sun within the Galactic plane has made understanding the
 true form of the Galaxy difficult. In the last ten years the situation has improved, however,  with the
advent of deep large areas surveys. Surveys
in the near IR such as 2MASS, UKIDSS or in the  visible such as the  Sloan Digital 
Sky Survey (SDSS, Newberg et al. 2002) have allowed a more precise modeling of the large scale
structures of our Galaxy.

It is generally accepted that there is both a thin and thick stellar disc. 
The stellar density of the thin disc is exponential in height above the Galactic plane and radial distance. 
In the vicinity of the Sun it has a scale length of about 2.4~kpc, however the scale 
height depends on the source type.  The older sources have a scale height of around 185 pc, but for the
younger sources it is considerably less than this.   
 The density of the thick disc (Gilmore \& Reid 1983) is also exponential  but with a larger
scale length and scale height (Bilir et al. 2008), although it has  less than a tenth
of the density of the thin disc on the plane in the vicinity of the Sun.  
   	 
 Of particular importance to our understanding of the Galaxy and  its formation, is to 
 determine what happens towards the outer edge of the disc.  Robin et al. (1992),  using visible
images from the CFHT, suggested that Galaxy has a cut 
of at galacto-centric radius ($R$) of 13.5~kpc when looking at $l=179^\circ$ $b=-2.5^\circ$,
 as the stellar density drops rapidly beyond that point.
L\'opez-Corredoira et al. (2002)  using 2MASS data  suggested that the drop in stellar density  near 
the plane was not caused by a cut off, 
rather a flare in the disc and concluded that the disc did not have a well defined cut off to at least 
15~kpc. 

The most recent deep surveys such as the  SDSS clearly show that there are 
stars out to at least $R$=20~kpc. 
Their distribution, however, is not consistent with a simple exponential disc particularly when 
looking  15 to 30  degrees from the plane. Some authors (Momany et al. 2006) suggest these 
can be explained by a flare of the disc.  
Many other authors, however, have preferred to attribute these sources to rings or streams  
beyond the edge of the disc. In particular, the sources being studied here have been attributed to the 
so called Monoceros ring/stream (Newberg et al. 2002; Rocha-Pinto et al. 2003; Conn et al. 2005, 2007, 2008).
This has been associated to the remnants of a dwarf galaxy which was cannibalized by the Milky Way,
and whose progenitor was associated to the Canis Major over-density (Martin et al. 2004;
Mart\'\i nez-Delgado et al. 2005; Bellazzini et al. 2006; Butler et al. 2007; de Jong et al. 2007;
Conn et al. 2007). This ring runs approximately parallel to the galactic plane, in the latitude range 
 $10<|b|<35$ deg. and over most of the second and third quadrants.

That stellar streams exist is well known, and that produced by the Sagittarius 
dwarf galaxy (Ibata et al. 1994)
is  probably the most well  studied stream associated with the Milky Way. Its  form and high angle  to the 
Galactic Plane makes  confusion
with a Galactic component highly unlikely.  However, the proposed  Monoceros stream is a very
different feature  to the Sagittarius stream running, as it does, parallel to the Galactic plane. 
Furthermore, whilst the  Sagittarius stream  is clearly a dwarf galaxy,
the progenitor of Monoceros in the over-density in Canis Major 
has also been questioned, as other authors have explained
the excess star as an effect of  the warped+flared 
disc of the Milky Way (Momany et al. 2004, 2006; 
L\'opez-Corredoira 2006; L\'opez-Corredoira et al. 2007).

In this paper we examine if, using a simple flared disc model as suggested by  Momany et al. (2006), 
we can reproduce the form of the deep star counts seen 15 to 30 degrees off the plane in 
Galactic anti-centre region, without requiring the presence of extra-galactic streams.

\section{The Method} 

  For this work we wish to count sources at a $R<$ 25~kpc towards the Galactic anti-centre
 and between 11 and 31 degrees off the plane. 

By far the simplest method of determining the  stellar density along a 
line of sight in the disc is by 
isolating a group of stars with the same colour and absolute magnitude 
within a colour magnitude diagram. 
This allows the luminosity function to
be replaced by a constant in the stellar statistics equation:

\begin{equation}
A(m)\equiv \frac{dN(m)}{dm}=\frac{ln\ 10}{5}\omega \rho
[r(m)]r(m)^3  
,\end{equation}\[
r(m)=10^{[m-M+5]/5}
,\]
where $\omega $ is the area of the solid angle in radians and $r$ is the
distance in parsecs; 
the differential star counts for each line of sight, $A(m)$, 
can be immediately converted into density $\rho (r)$. 

In the near IR, the red clump giants have been successfully used as 
standard candles, particularly when looking in the inner Galaxy 
(L\'opez-Corredoira et al. 2002).  
However,  the red clump stars in the outer disc at the distance of interest would appear 
at $m_k$=15 and so the local dwarfs with the same colour would completely dominate the counts 
 at this magnitude 
(L\'opez-Corredoira et al. 2002). 
 
 As the areas of
interest all lie well off the plane, then the extinction will be low (typically under 1 magnitude). 
Therefore, it is possible to use visible counts.  An examination of the  HR diagram shows that when 
the extinction is low, the late F and early G dwarfs can be isolated using colour with only minimal 
contamination from other sources with the same colour but different
absolute magnitudes (sub dwarfs,galacto-centric radius  giants etc). For this work we have selected the 
source between F8V and G5V. 
This gives a range of $g-r$ of 0.36 to 0.49
and a range of absolute magnitudes $M_g$=4.2 to 5.4 which  makes the sources approximately   $m_g\sim 20$ 
at the distance of interest.  By having a range of absolute magnitudes, then when the counts are converted to 
density vs distance
there is  some smoothing and this is not included in a model which assumes a single absolute magnitude.  
Although the selection used here is not as well defined as the red clump stars in the near 
IR, it still has a sufficiently small range in absolute magnitude that the smoothing has little effect  
and the above approximation 
remains valid. There will, however, 
be sufficient stars detected in the outer Galaxy to give meaningful statistics which would not be the case 
if a smaller 
range in absolute magnitudes were used. Earlier sources have 
not been included as these sources would belong to a younger population with a 
far smaller scale height, also the absolute magnitude changes
far more rapidly with colour. Later sources would have significant giant contamination and again the absolute 
magnitude changes more rapidly with colour.

The areas selected for this work lie between 10 and 30 degrees from the plane and are within 40 degrees of
the anti-centre.  We have not extended the  range of longitudes because of the Galactic warp. 
The warp means that, in effect the position of the Galactic plane varies 
with position in the Galaxy, and  along the line of sight.  Although the effect of warp is strongest towards 
the  outer edge of the 
disc, where the plane can be 
over a kpc away from the expected position with no warp, it can be clearly
seen in the counts within a few kpc of the Sun (Hammersley et al 1995, L\'opez-Corredoira et al. 2002, 
Momany et al. 2006). 
The models of the warp are normally simple, based on  tilted rings, and in general  these do provide 
a good representation of the star counts. However, 
in the very outer disc the models are harder to test, and  a small error  
in the model  can lead to a significant error to the predicted position of the plane, 
and hence the predicted  densities. 
Therefore, we have limited the regions used for this work to those  
where the effect of the warp should be small 
and so can be ignored. It should be noted, however, that whilst it will be possible  to ascertain whether 
a flare can reproduce the star counts seen in the outer disc, it will not be possible to accurately determine 
all of the parameters.  When large scale deep star counts in the outer disc become available  it will 
be possible to  model the warp more accurately. Until then, however, great care must be taken when 
comparing models with counts in the outer disc.     
   
  A second reason for limiting the longitude is that the distance to the feature increases  rapidly  with distance from 
  the anti-centre,  making  
 the sources fainter  and appear  closer to the plane so that there is less contrast with the other disc sources.

     \begin{table}
      \caption{Lines of sight used.}
         \label{tab1}
	$
         \begin{array}{llll}
            \hline
            \noalign{\smallskip}
            l(^\circ ) & b(^\circ ) & Area\ (deg^2) & Extinction \ E(B-V)\\
            \noalign{\smallskip}
            \hline
            \noalign{\smallskip}
	       150   &   15   & 1.94  & 0.25  \\
	       180   &   11   &  0.45 & 0.241 \\
               183   &   21   &  1.22 & 0.054 \\
               183   &   31   &  3.27 & 0.052 \\
	       223   &   20   &  1.00 & 0.035 \\
            \noalign{\smallskip}
            \hline
	    \vspace{0.5cm}
         \end{array}$
   \end{table}

   \begin{figure}
   \centering
   \includegraphics[width=8cm]{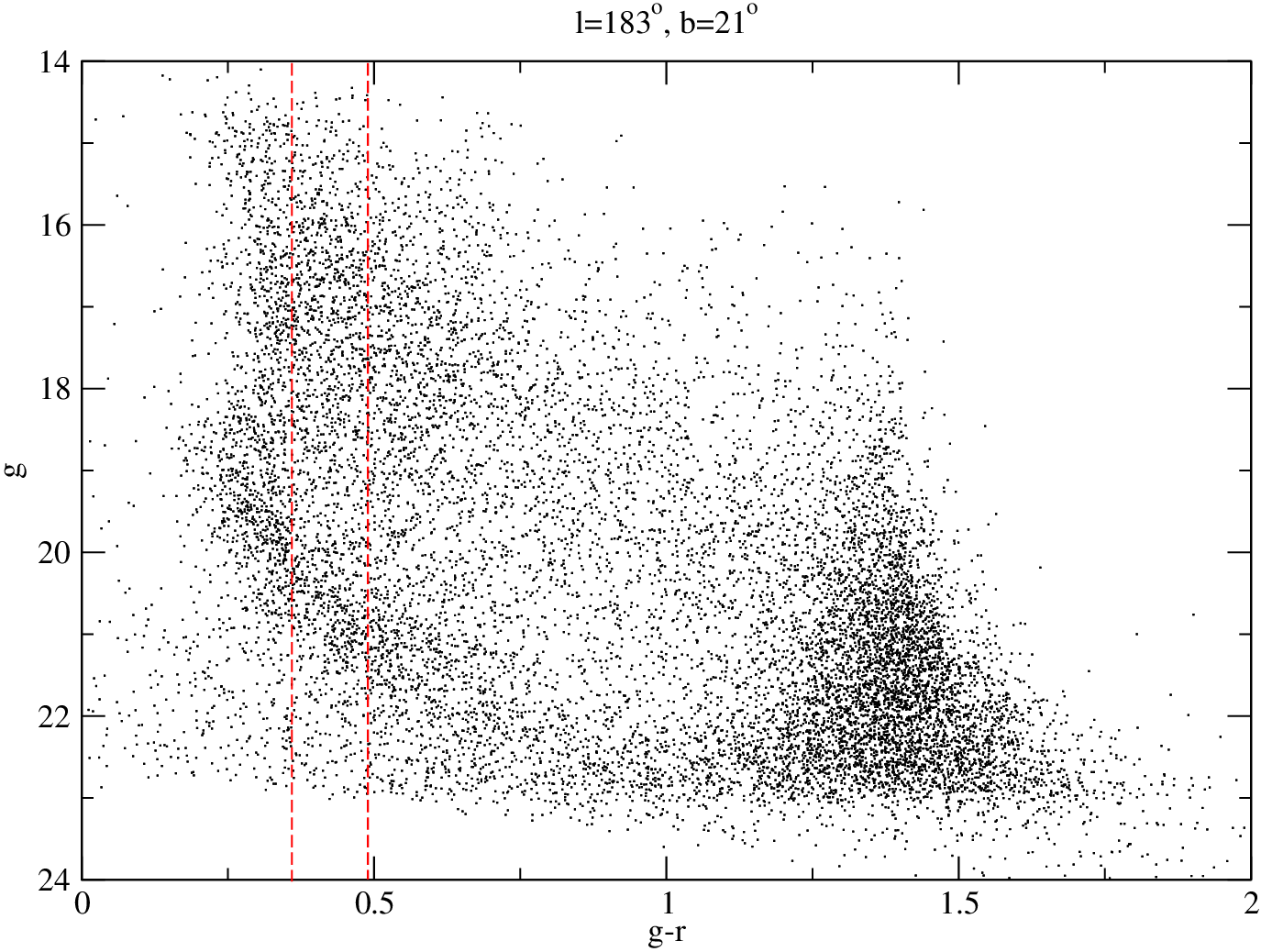}
   \caption{The CMD  for the region  $l=183^\circ $, $b=21^\circ$  }
   \label{Fig1}
   \end{figure}

\section{The Data} 
 The source densities in these regions will be very  low requiring square degrees of sky to be 
  covered to provide sufficient counts to give reasonable statistics. Therefore, the data has been taken from 
 the SDSS release  DR7 (Abazajian et al. 2009) with regions $\sim 1$ square degree. 
 As the regions are all well off the plane it has been assumed that 
 the extinction is local and so all of the magnitudes have been corrected for extinction using the 
 Galactic extinction model of Shlegel et al (1998). 
 Furthermore, the extinction is relatively small and a small residual error will not 
 significantly affect the results, although it
 would make the sources appear at the wrong distance.  
Table 1 shows the position and assumed extinctions. The positions $l=183^\circ$, $b=21^\circ $ and
$l=220^\circ $ where used by Newberg et al. (2002); and the position at $l=150^\circ $ 
was used by Conn et al. (2005), to support their ``Monoceros stream'' hypothesis.

   \begin{figure*}
   \centering
\includegraphics[width=6cm]{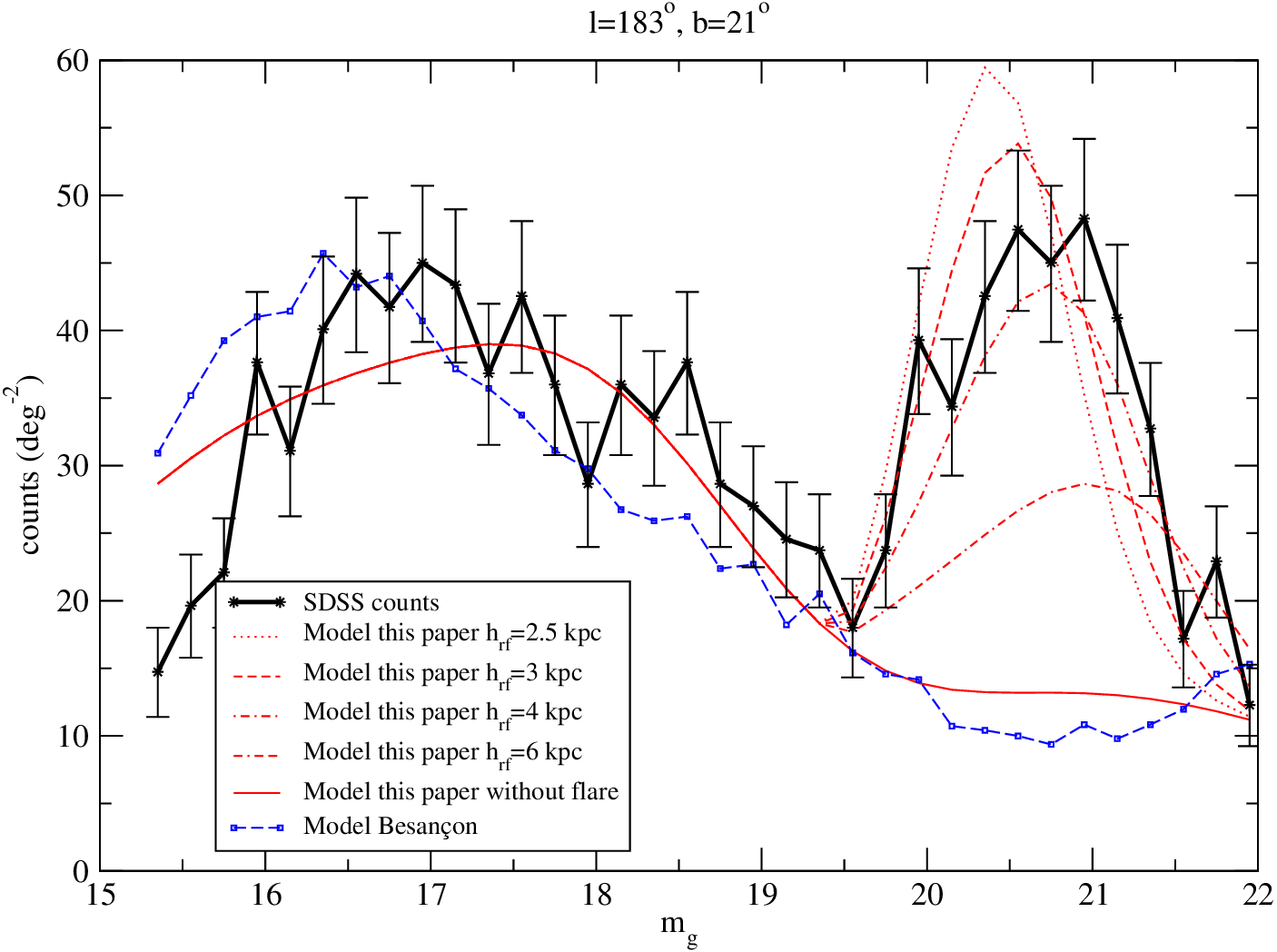}  \hspace{0.5cm}
\includegraphics[width=6cm]{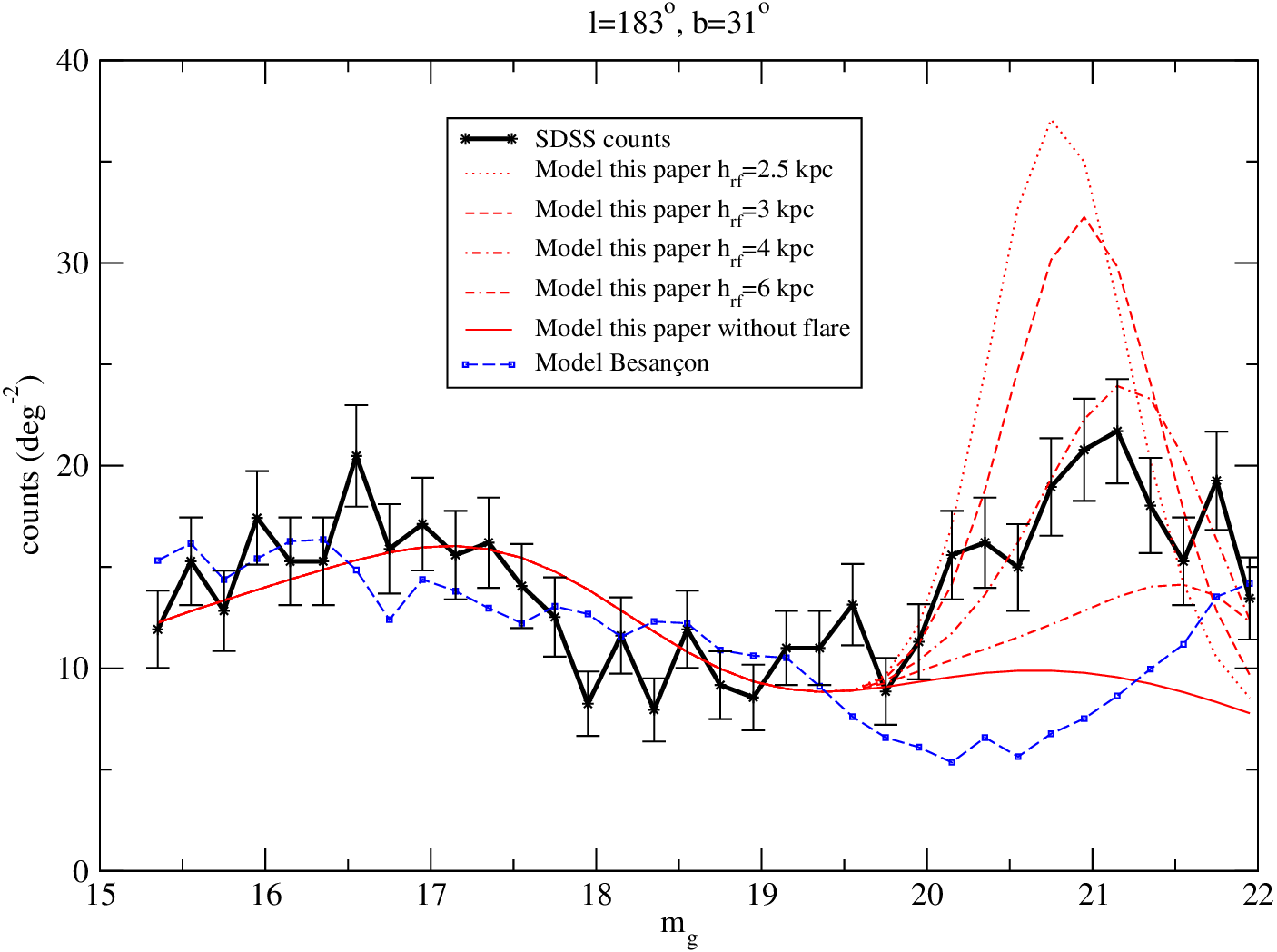} \\
\vspace{0.75cm}
\includegraphics[width=6cm]{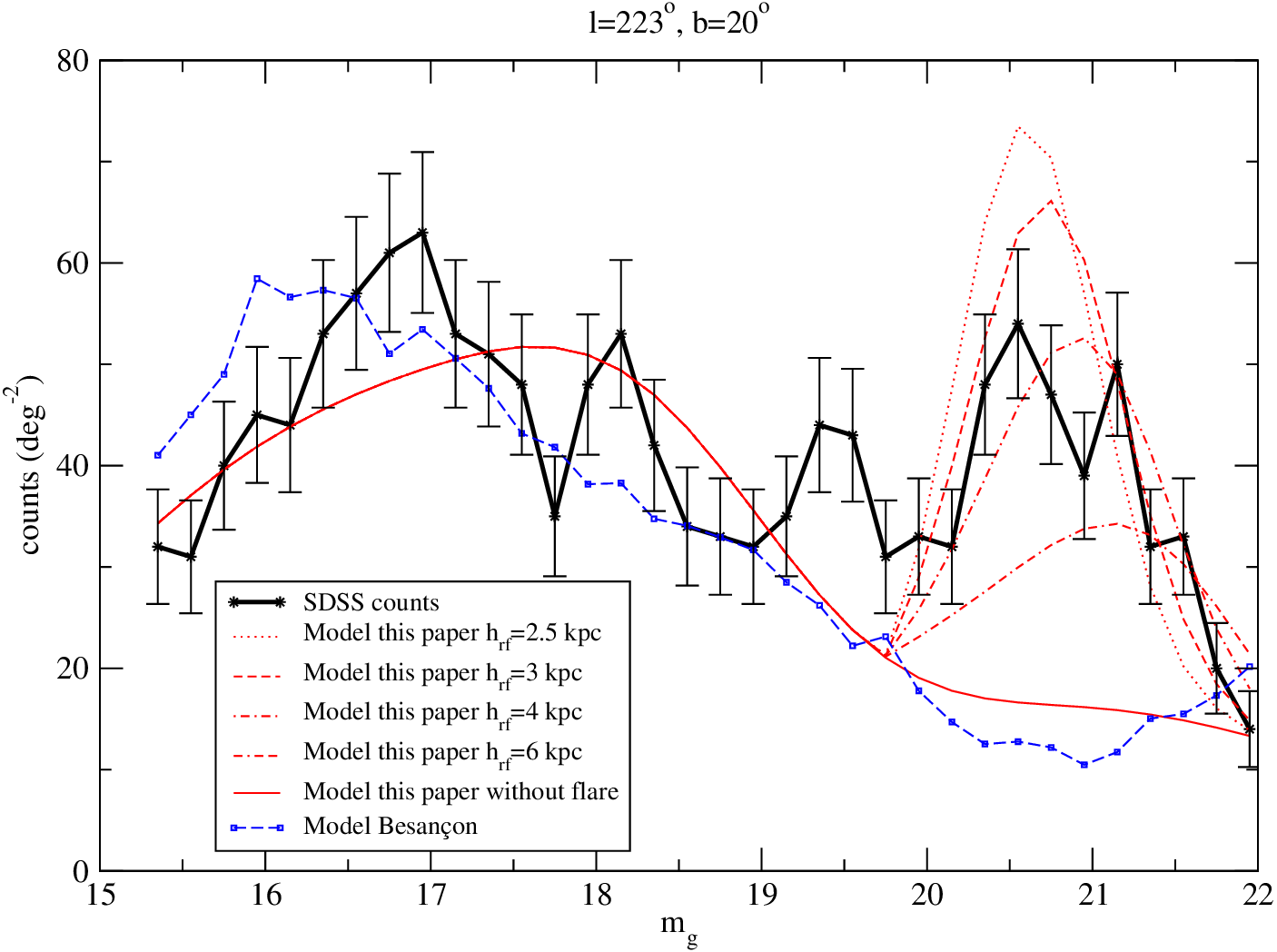}\hspace{0.5cm}
\includegraphics[width=6cm]{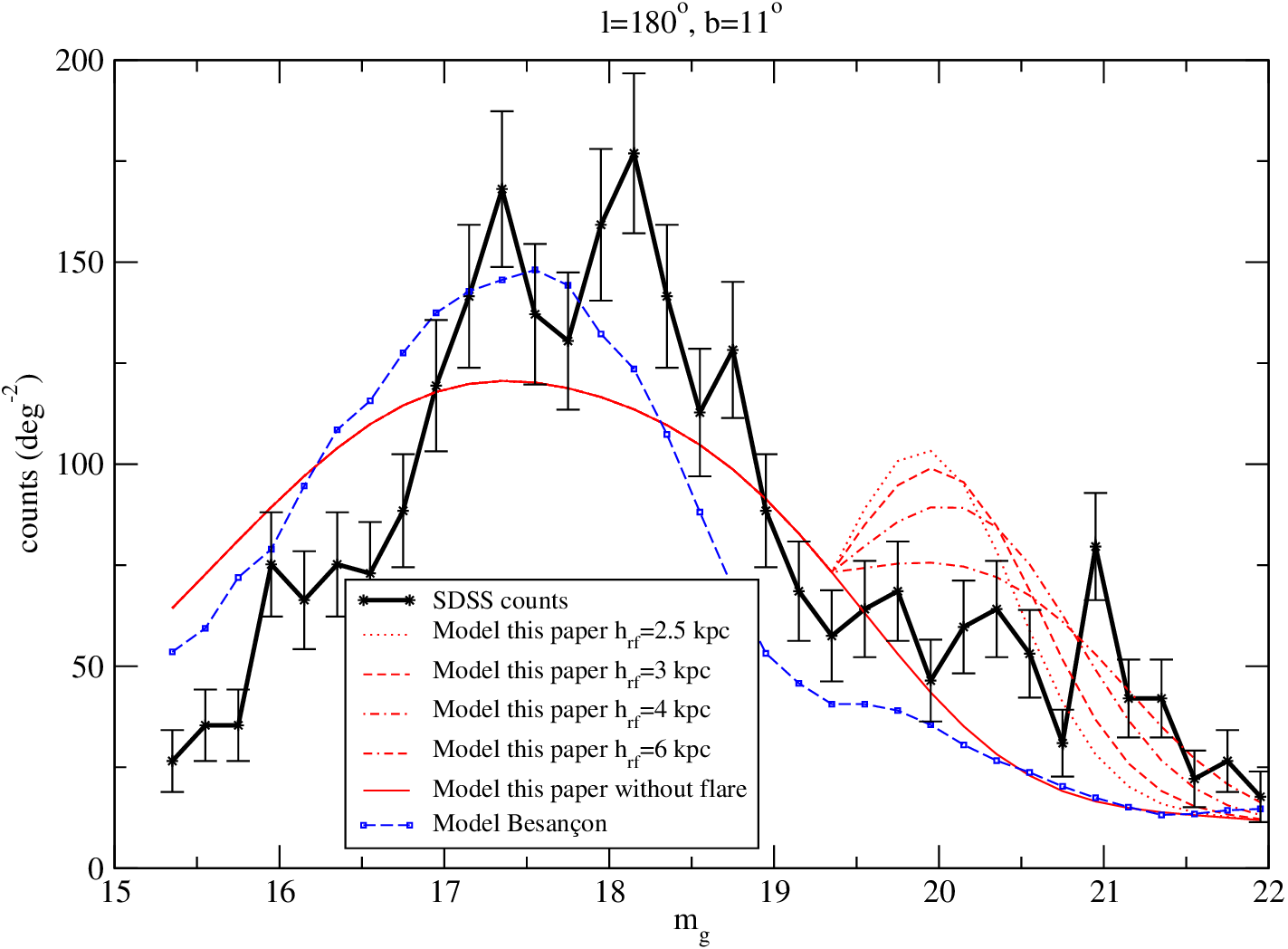} \\
\vspace{0.75cm}
\includegraphics[width=6cm]{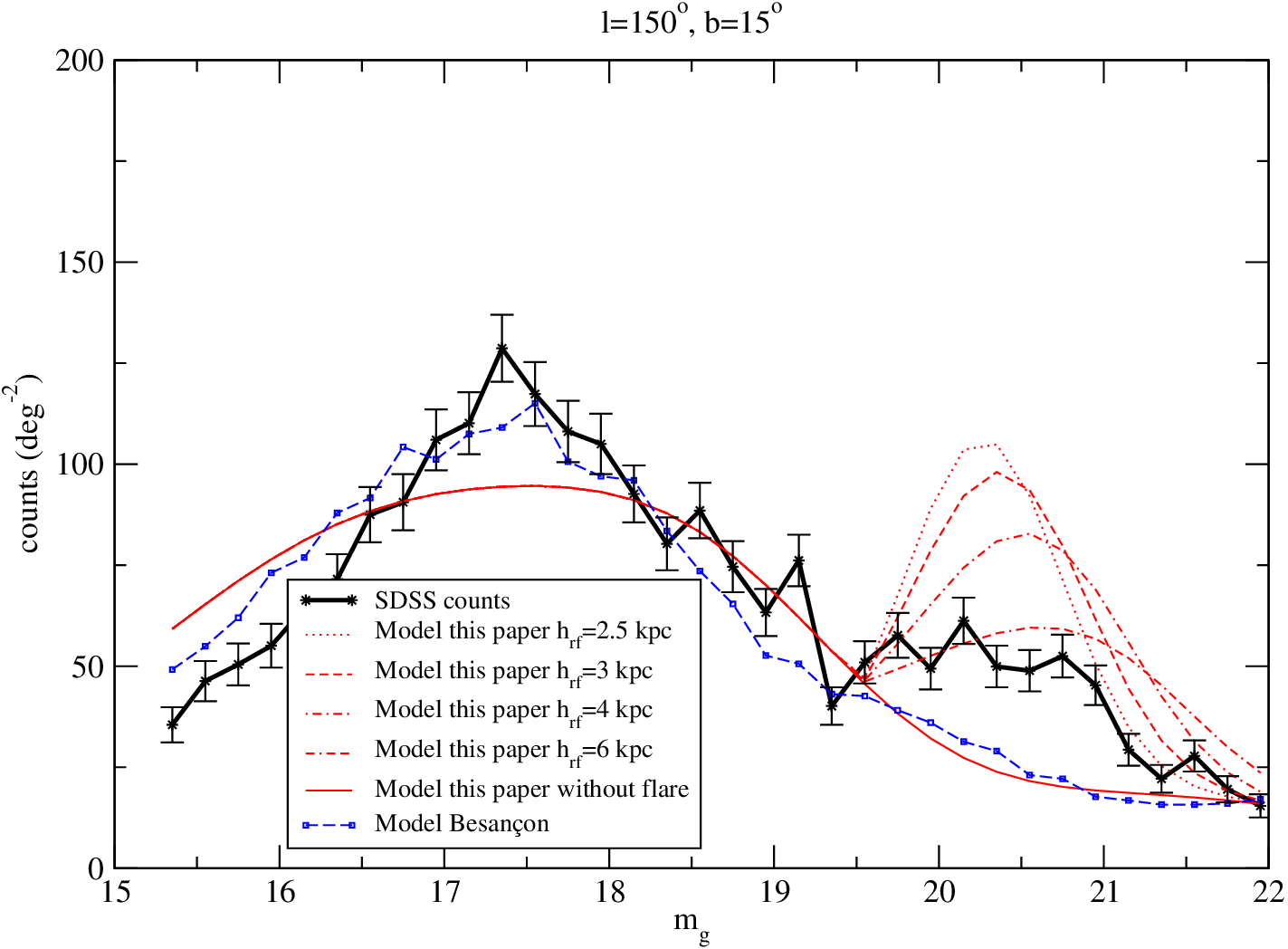}
\caption{The differential star counts with the error bars  in  five regions towards the anti centre. As well as the counts, the predictions of  
the Besan\c{c}on model and a
simple flared  or non-flared disc model are also shown.}
   \label{Fig2}
   \end{figure*}

   \begin{figure*}
   \centering
     \includegraphics[width=6cm]{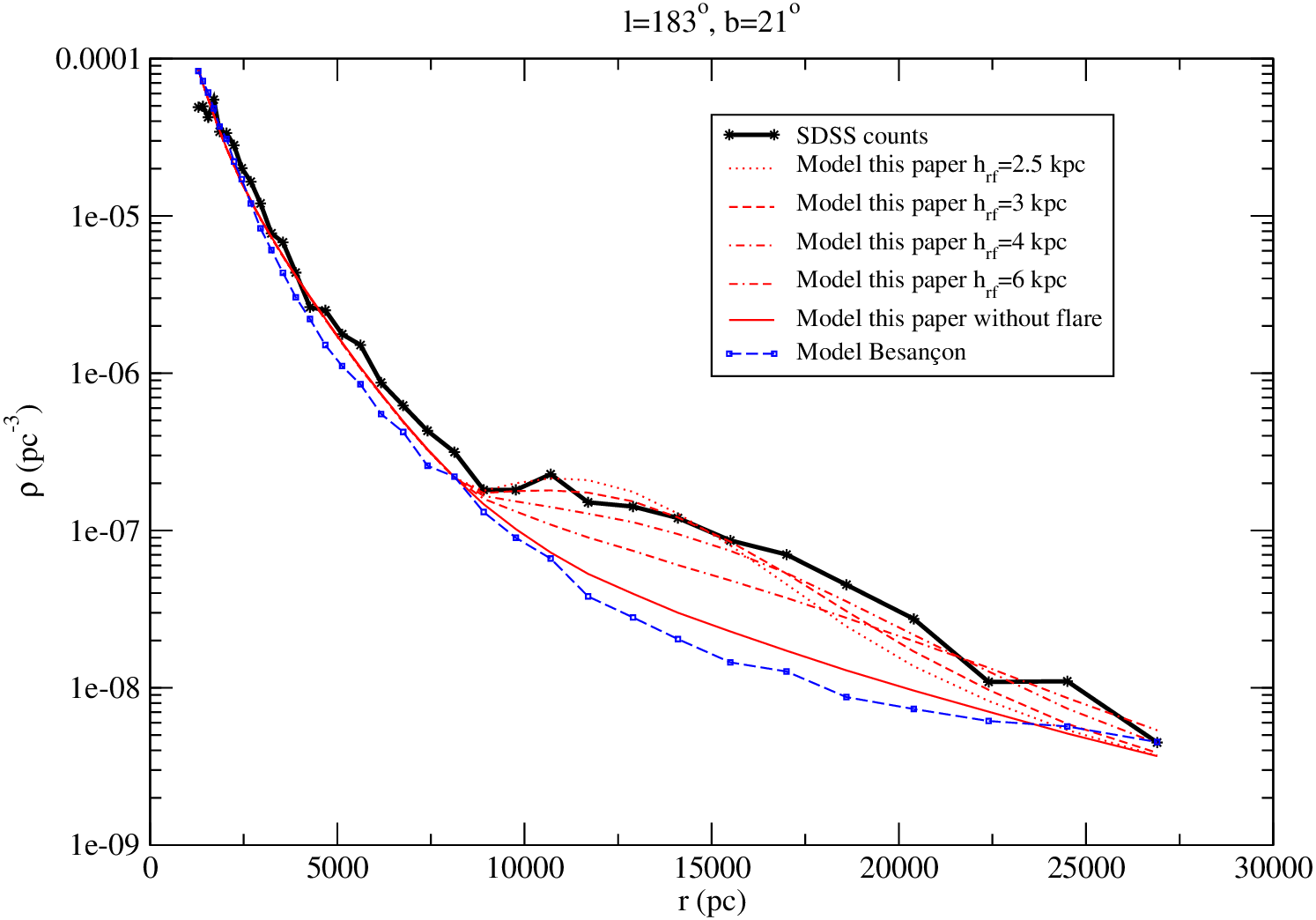} \hspace{0.5cm}
     \includegraphics[width=6cm]{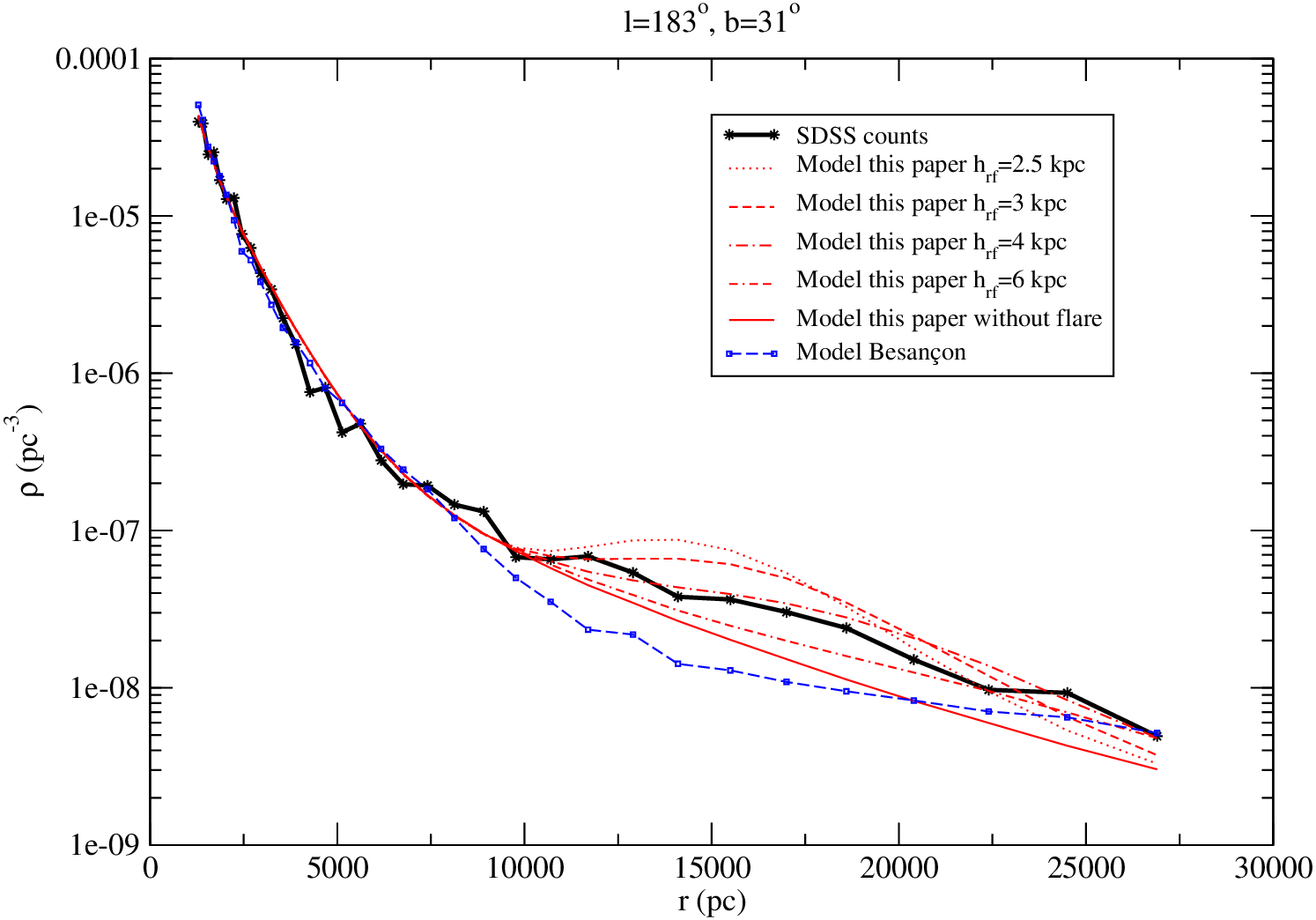} \\
      \vspace{0.75cm}
   \includegraphics[width=6cm]{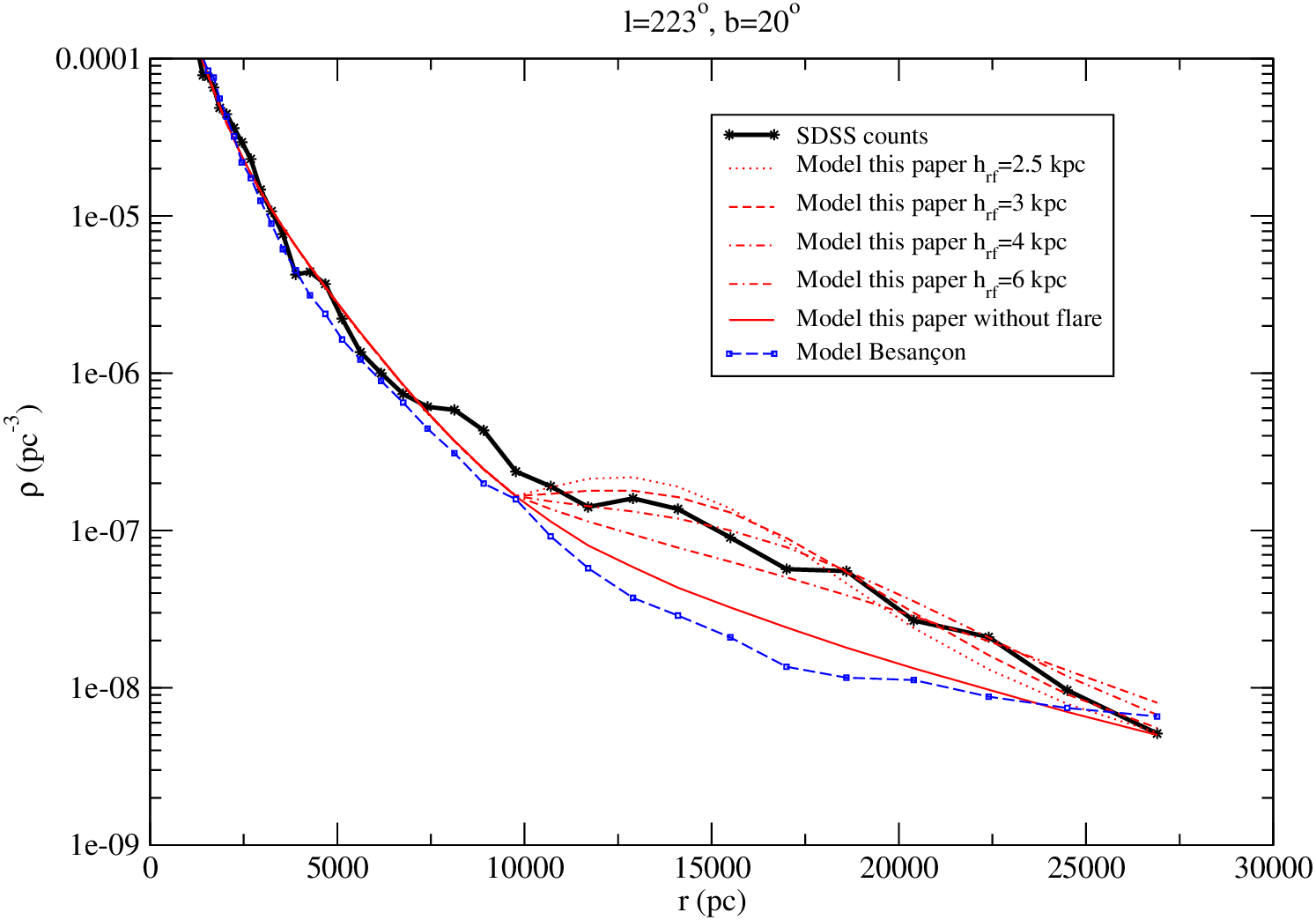} \hspace{0.5cm}
   \includegraphics[width=6cm]{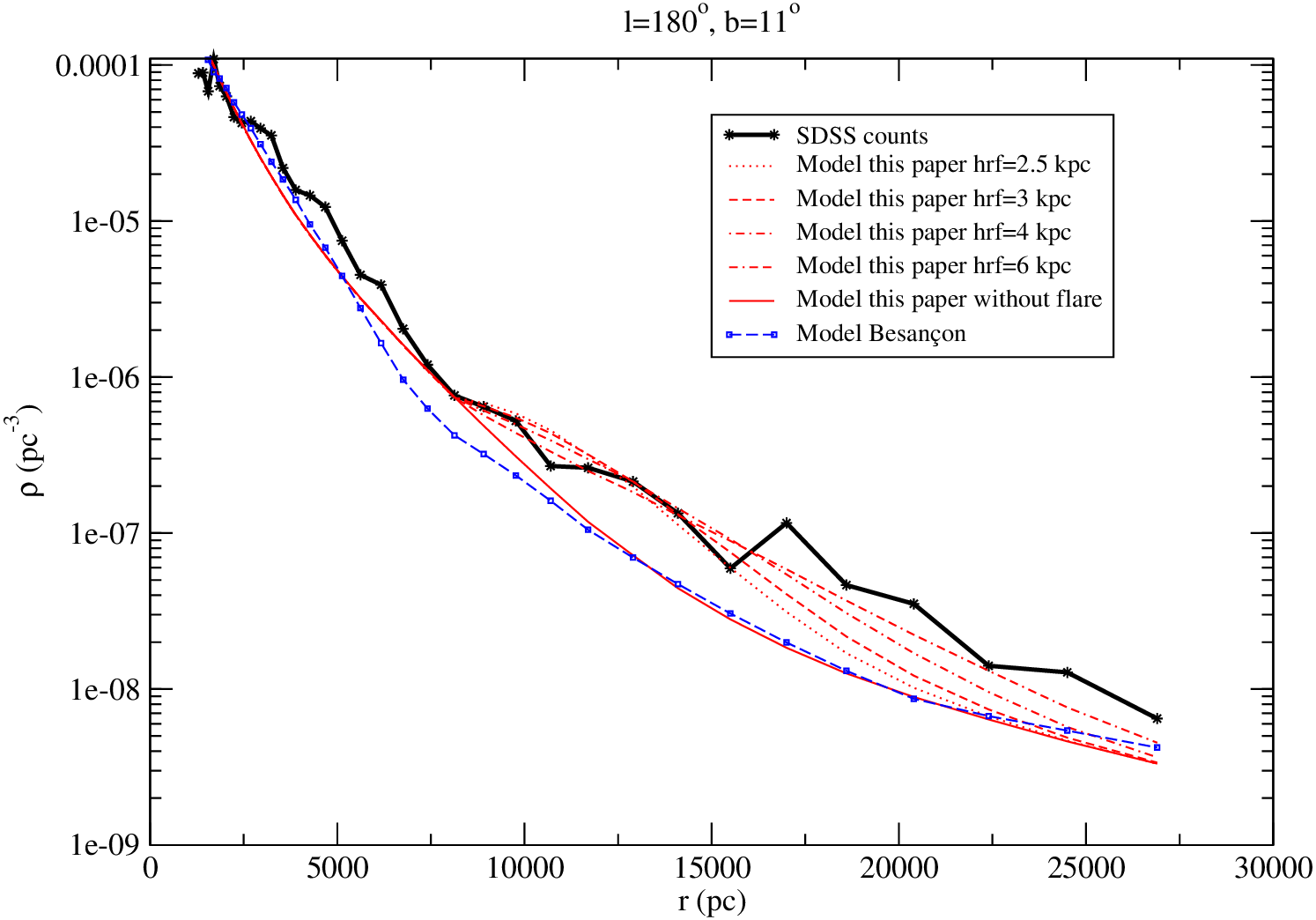} \\
         \vspace{0.75cm}
\includegraphics[width=6cm]{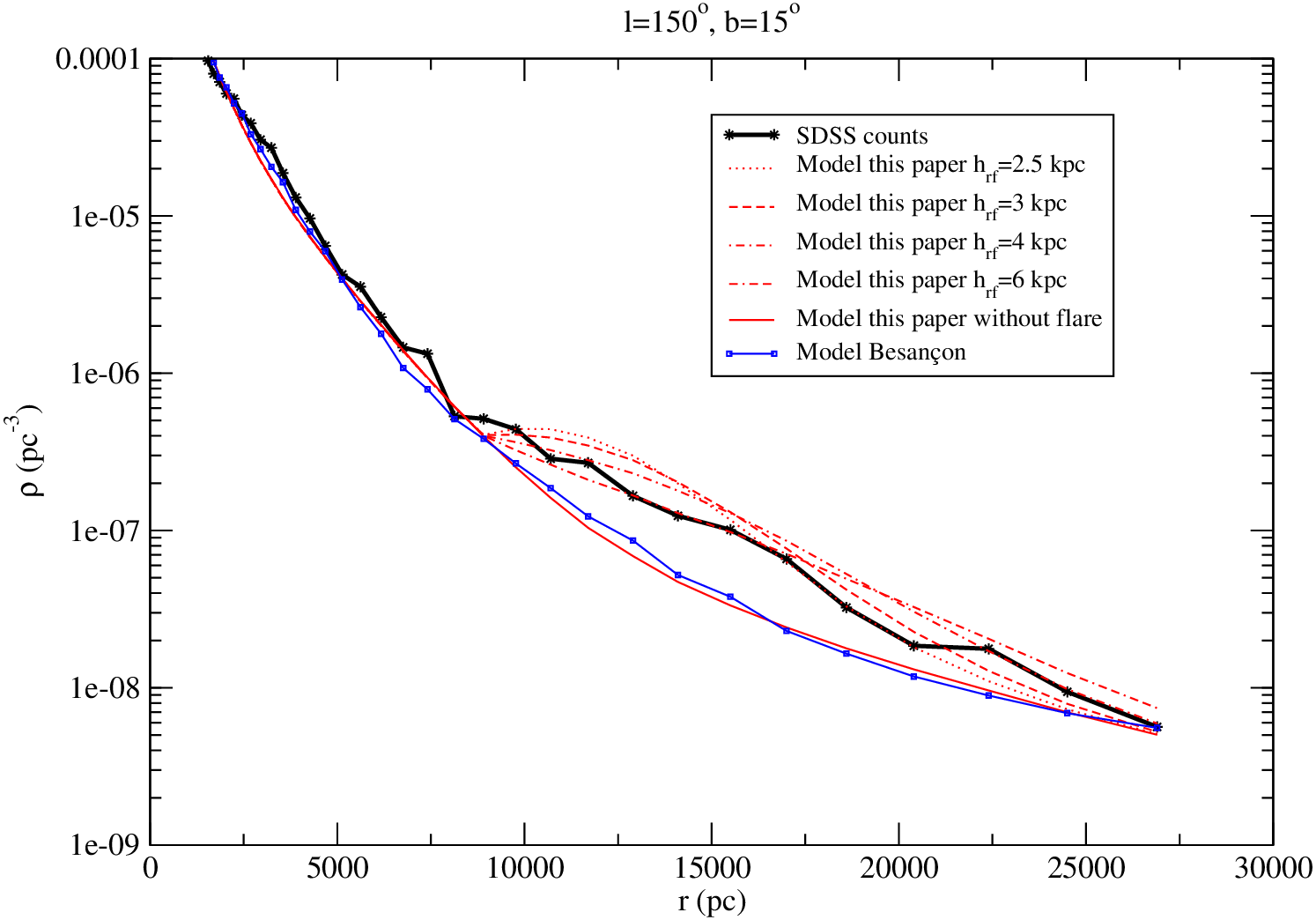} 

\caption{The stellar density with R along  five lines 
of sight towards the anti centre.As well as the counts the predictions 
of the  Besan\c{c}on model 
and a simple flared  or non-flared disc model are also shown.}
   \label{Fig3}
   \end{figure*}

 \section{ Results} 
 Figure 1 shows the  $g$ vs. $g-r$ colour magnitude diagram  for the region  $l=183^\circ $, $b=21^\circ$
with the sources of   interest marked. It can be
 seen they are in a reasonably uncluttered part of the diagram
 allowing them  to be isolated. Furthermore, there are no 'holes' in the area of interest 
 indicting that there is no 
 areas of high extinction.  
  Figure 2
 shows the  differential counts for the selected colour  along with the 
 Besan\c{c}on model (Robin et al. 2003) for these stars. This model is widely used 
 and generally accurately predicts the counts over wide areas of the sky.  
Here we have only run the model for
 the stars between F9V and G4V. The model was run with no extinction and predicted 
counts in $V$  which were then  converted to $g$ using the $g-V$ colour for these stars. 
 The  Poissonian error bars  from the counts in each magnitude bin is also shown.

  The counts have
 two clear peaks, one at about $m_g=17$ and a second at $m_g=$20.7 with a minimum at 
 $m_g=19.5$  (see Fig. 2). This is seen most clearly at  $b=21^\circ $, and  $b=31^\circ $, 
 however even in the region $l=180^\circ $, $b=11^\circ $  there is  some
 evidence for excess counts.    
 The Besan\c{c}on  model does give a
 good fit to the data for $m_g<18$, however, at $m_g=20$ there  is a sharp rise in the 
 counts which is not predicted by the model. The peak corresponds with a
 distance of about 15~kpc from the Sun or a $R=23$ kpc.
 This second peak cannot be caused by the halo, which is included in the model Besan\c{c}on 
 as the peak for  these counts is predicted to start at $m_g\approx 23$ and reach a peak at about $m_g=25$.   
 Therefore, there are two possible alternatives;  this is part of the disc or 
 an extra galactic component, such as a stream beyond the disc of the Galaxy.
  
 \section{Analysis}

These star counts cannot be produced by  a simple exponential disc, even when including both 
the thin and thick disc, as such a model cannot produce this second bump (see figure 2). 
Even if there were no cut off this would only slightly affect the counts.    
However, a flared disc with no cut off  may produce the 
second peak along specific lines of sight, as it would on average distribute stars away  from the 
plane. In order to explore this  possibility,  
 Eq. (1) has been used directly determine  the stellar density 
 with distance assuming that the luminosity function
 can be replaced by a delta function with   $M_g=4.80$.  
 Figure 3 shows the density with galacto-centric radius for the lines of 
 sight used. It is notable that the  stellar density is
 continuously decreasing, although it does change slope,  without significant gaps or bumps suggesting 
 that it may be explained only one continuous Galactic structure 
 and not two separate structures. Hence  the second bump in Fig. 2  is an effect of the shallower slope  
 in the stellar density 
being multiplied by the factor 
 $r^3$ [see Eq. (1)]. 
   
In order to determine the effect of a flare, a simple model was built containing
a thin disc(scale height $h_{z,thin,\odot }=186$ pc and 
 scale length $h_{R,thin}=2400$ pc) and a 
 thick disc (scale height $h_{z,thick,\odot }=631$ pc and scale length 
 $h_{R,thick}=3500$ pc) 
 were assumed with the density of the thick disc being 9\%
 of that of the thin dick on the plane at the solar radius
 (as observed at high latitude with SDSS; Bilir et al. 2008).
  The halo counts of the model by Bilir et al. (2008) were also included, although
 they give a negligible contribution at these magnitudes.
 It was assumed that the flare increase  the scale height
 of both the thin and thick discs  beyond a certain
 galacto-centric radius, and that the size of the flare increased exponentially.  
 It should be noted, however, that the thin disc 
 only provides a significant contribution to the counts  where the line of sight is within a few 
 hundred pc of the plane.  
  The flare begins at 
 about $R=16$kpc by which distance the lines of sight are over a kpc from the plane and hence the contribution 
 of the thin disc to the counts where the flare is important is negligible. Therefore, this paper is 
 only analysing the effect of the flare on the thick disc. Lines of sight far 
 closer to the plane would be required in order to analyse the effect of the flare on 
 the thin disc, and only then can any relationship
  between the flare in the thin and thick discs determined.   
 
  It was also assumed that 
  the  total number of 
 stars at a galacto-centric radius followed 
 a simple exponential model. If this is not done then the large scale height caused by the flare 
 would lead to there being more stars 
 in total at that galacto-centric radius. 
  
 The variation of stellar density is given by
 
 \begin{equation}
 \rho=\rho_{thin}+\rho_{thick}+\rho_{halo}
 ,\end{equation}
 \[\rho_{thin}=
 A\left(\frac{h_{z,thin,\odot }}{h_{z,thin}(R)}\right)
 \exp\left(-\frac{R-R_\odot}{h_{R,thin}}\right)\exp\left(-\frac{|z|}{h_{z,thin}(R)}
 \right)
\]\[
 \rho_{thick}=
 0.09A\left(\frac{h_{z,thick,\odot}}{h_{z,thick}(R)}\right)
 \exp\left(-\frac{R-R_\odot}{h_{R,thick}}
 \right)\exp\left(-\frac{|z|}{h_{z,thick}(R)}\right)
 \]\[
 \rho_{halo}=1.4\times
10^{-3}A\frac{\exp\left[10.093\left(1-\left(\frac{R_{sp}}
{R_\odot}\right)^{1/4}\right)\right]}{(R_{sp}/R_\odot)^{7/8}};
 \]\[
 h_{z,thin/thick}(R)=
 \left \{ \begin{array}{ll}
        h_{z,thin/thick,\odot },& \mbox{ $R\le 16$ kpc} \\
        h_{z,thin/thick,\odot }\exp\left(\frac{R-16\ {\rm kpc}}{h_{rf}}\right),
	& \mbox{ $R>16$ kpc}
\end{array}
\right \}
\]\[
 R_{sp}=\sqrt{R^2+2.52z^2}
\]

 This is, therefore, the standard thin and thick exponential disc model, but with a flare starting at a specific
galacto-centric radius. There are  only two free parameters, the position of the start of the flare and the scale length 
of the flare  
 
 It should be noted that the aim is not to try and fit the model to the data but to  demonstrate that a simple model 
 does reproduce the counts.  Effects such as errors in the extinction calculation, residual effects or the warp and the 
 expected none axi-sysmetry of the outer disc would be require a far more complex model and a far larger coverage of sky,
 in particular closer to the plane.   
 
 In Figures 2 and 3 the effects of various flares  starting at a 
 galacto-centric radius of 16~kpc are shown. For Fig. 2 the amplitude $A$ 
 has been normalised  to 
 the measured counts between $m_g$ of 16 and 19, which is well well before 
 the flare starts. In Fig. 3, however,   the density  has been normalized  to the measured
 $\int _0^\infty \rho (r)dr$. 
 
 Fig. 4 takes 
the densities and determines by how much the disc would need to be 
flared at each galacto-centric radius to give the determined density. At the radius of the Sun 
the relative scale height is 1 and it remains constant until 
 a $R$ of 16 kpc, at which point the  scale height increases rapidly. 
 It is true that the flare is not identical in each case, however they are 
very similar with a typical flare   scale length of 4.5 $\pm$ 1.5  kpc (Fig 4). There are, however, 
a number of effects  which have not been taken into account.
\begin{itemize}
\item The  warp has been ignored here, but  would mean that the average plane position would be above
$b=0$ in the in the  $l=150^\circ$ region and below $b=0$ at $220^\circ$. In the region 10-18 degrees from the plane  the counts
the flare increases rapidly, and so a small error in the position of the plane  would  significantly 
reduce the counts at $l=150^\circ$ $b=15$ . 
\item  The simple model
used here assumes a circularly symmetric outer Galaxy. As here we are dealing with the thick disc at 6 to 7 scale lengths
from the Galactic centre, 
then if there were a  10 \% 
change in scale length with Galactic longitude it would lead to a 50\% change in counts.
\item The model used here assumes that the parameters of the flare do not depend on Galatic longitude.
\item  The extinction model used here is very simplistic. 
\end{itemize}
Finally it should be noted that the counts for most distant  points are very low leading to a large scatter.
 When  far more regions become available it will be possible to quantify these effects and so refine the results.

These results do show that the variation of the number of stars with galacto-centric 
distance is consistent with an exponential law out to over 20~kpc.  There is also no evidence for either a major drop in counts 
just before the flare starts or a sudden jump in the counts. This means that the counts in the  disc can be described by a single 
function to beyond 20 kpc.    
 
These plots show that the simple flare model not only predicts that at 20 degrees from the plane 
{bf that there should be two peaks in the star counts}, but
can  reproduce the numbers of counts and form.

  The alternative scenario is a ring of stars centred at $R$= 23~kpc, which
is  detected for  $R>16-17$ kpc (Conn et al. 2007). 
 As this structure would only exists in the outer Galaxy, its parameters are not constrained to fit
observations in the solar neighbourhood or elsewhere in the Galaxy. This means that 
there is a wide range of possible parameter space that can be explored and hence 
by adjusting the width and density, a ring can be made to fit second peak. However, 
this ring must start  
very
close to the edge of the disc.  If the inner part of the stream is more than  1 or 2~kpc  from the edge of the disc 
then  there would
be a  noticeable drop in the stellar density at that point. Conversely if it overlapped with the edge  of the disc 
then there would be a
jump in the density. Furthermore, the number of stars  in the stream  has to be very  similar to those predicted 
for the disc with no cut off, otherwise none of the flared models  would come close to  reproducing  the 
measured counts. Finally 
the vertical distribution of the counts again has to mimic the flared disc model, and this is over more that 20 degrees
in latitude which correspond to 3 to 4~kpc at an $R$ of 15 to 20~kpc.  Hence, if there is a  stream  it would have to have
many characteristics which are identical to those predicted for a flared outer disc.

\section{Discussion }

That the disc of the Galaxy is flared is widely accepted and is particularly
well studied in the HI  (e.g., Kalberla et al. 2007). 
They also find that the flare becomes relevant at around $R$=16~kpc and
  the disc continues to at least 23~kpc.  There may be also a dependence of the scale height with the 
galacto centric azimuth, as observed in the gas too (Kalberla et al. 2007), and this
lopsidedness could find an explanation on the non-axisymmetrical distribution of pressures
in the outer disc (L\'opez-Corredoira \& Betancort-Rijo 2009)

By, in effect, moving sources 
away from the plane the result is that once the flare starts there will be a
sharp drop in the counts when compared to the predictions of a simple
exponential disc for any lines of sight 
near the plane, whereas at higher latitudes there will be considerably more
stars than predicted.    The sudden loss of stars beyond about $R$=14~kpc is reported in
Robin et al. (1992) when looking at $l=179^\circ$ $b=-2.5^\circ$, and the results shown here shows
that the second peak in the differential star counts at $b=20^\circ $ is entirely 
consistent with a flare.  

  The stars that are in the second peak would belong to the thick disc as they are typically  
 3 to 4~kpc above the plane and are at two to three times the galacto-centric radius of  the Sun.
Their metallicity has been measured to be an average of [Fe/H]=-0.96, with an rms
scatter of only $\sim 0.15$ dex (Ivezi\'c et al. 2008), which is compatible with the metallicity
of the outer thick dick (L\'opez-Corredoira et al. 2007, Fig. 3).

 The flare of the thick disc is important for understanding its origin, as current theories suggest 
 that thick discs caused by mergers will always flare. Bournaud et al. (2009) 
 found  that
if a flare were caused by minor mergers then there would be no  
region of the thick disc with a constant scale height. The analysis 
presented  here is  not  
sensitive to changes in the scaleheight of thick disc  closer to the solar circle as the
counts  for $R<16kpc$ along the lines of site used  
 are dominated by the thin disc.
 An analysis of the thick disc flare in the whole range of  
galactocentric distances would be necessary to see whether 
the mechanism proposed by Bournaud et  
al. (2009) is applicable here.

   \begin{figure}
   \centering
   \includegraphics[width=8cm]{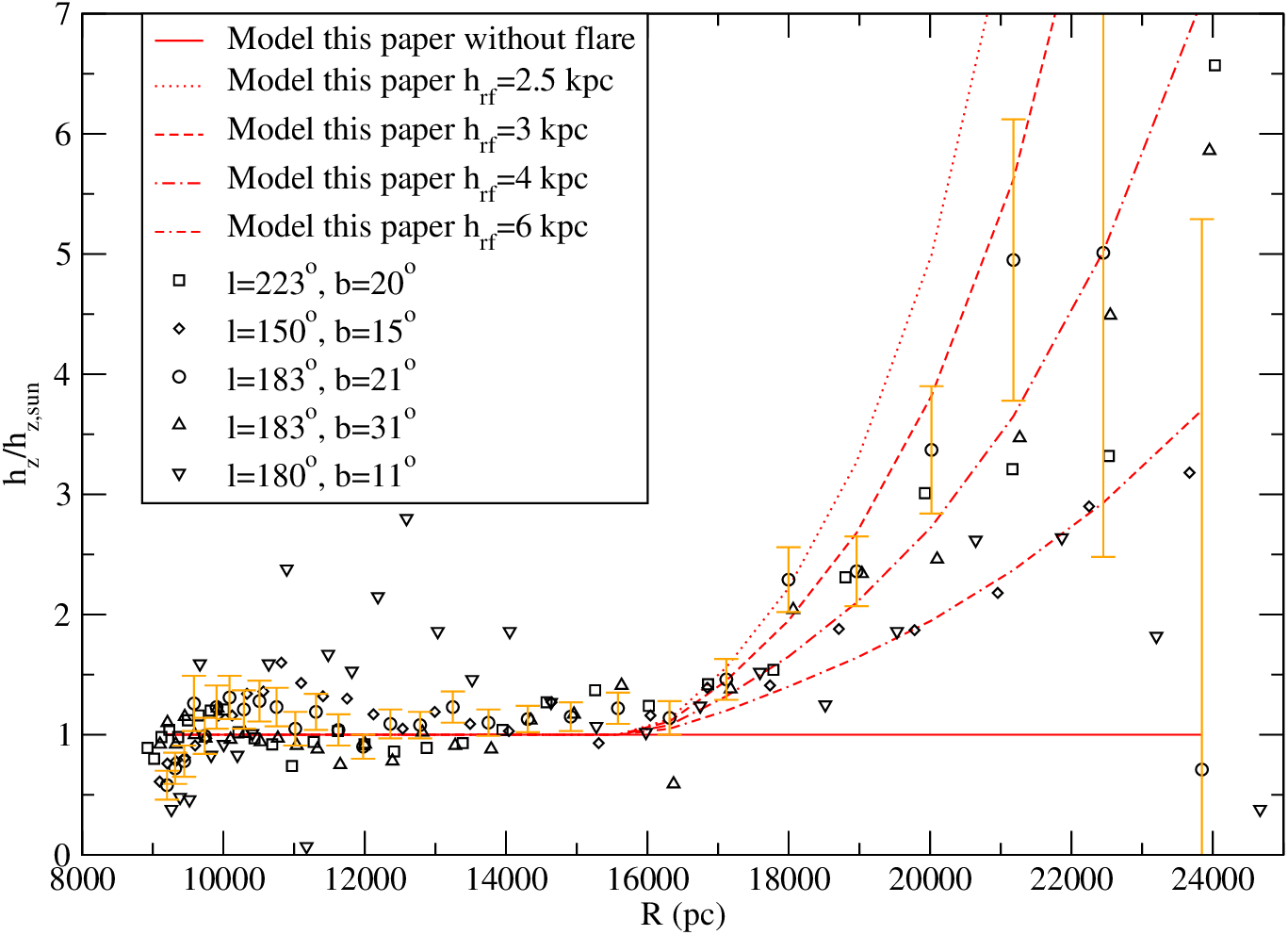}
   \caption{The proportional flare in the scale height required to give 
   the counts at each magnitude along each line of sight.  The error bars from the number of sources 
   detected in each magnitude bin are shown for one region.}
   \label{Fig4}
   \end{figure}

\section{Conclusions}
Uncertainties in
our knowledge of the disc makes asserting the presence of the Monoceros stream 
principally because  a model does not accurately reproduce the star counts dangerous, 
even when using a model as good as the Besan\c{c}on model. If the disc has a 
cut off at  $R$ of about 14~kpc, as used in the Besan\c{c}on model, then the 
ring would be necessary to give the star counts detected.  If, however, there is no cut off 
and the thick disc instead flares, then the bulk of the stars in the regions  being discussed here cannot belong to a stream 
as there are only sufficient stars to be explained by the disc alone. It should be noted that this would not preclude there
being smaller scale streams towards the anti-centre affecting smaller regions of sky.

 We have shown that a flare in the thick disc  would leave a very clear 
 signature in the star counts 
in the outer Galaxy, and that this signature is seen.    
We have shown that both the form and number of stars detected towards 
the anti-centre between 11 and 31  degrees from the plane can be readily explained by a  
flared thick disc without a cut-off. When a larger area of sky becomes available it will be possible to refine 
the outer disc and flare models and so improve the quality of the fit, in particular nearer the plane. Hence, 
the presence of the extra-galactic  
Monoceros ring/stream is not required to explain the star counts.

\begin{acknowledgements}
 MLC was supported by the {\it Ram\'on y Cajal} Program of the Spanish Science
Ministry. 
Funding for the SDSS and SDSS-II has been provided by the
Alfred P. Sloan Foundation, the Participating Institutions, 
the National Science Foundation, the U.S. Department of Energy, 
the National Aeronautics and Space Administration, the Japanese 
Monbukagakusho, the Max Planck Society, and the Higher Education 
Funding Council for England. The SDSS Web Site is http://www.sdss.org/.
The SDSS is managed by the Astrophysical Research Consortium 
for the Participating Institutions. The Participating Institutions are the 
American Museum of Natural History, Astrophysical Institute Potsdam, 
University of Basel, University of Cambridge, Case Western Reserve 
University, University of Chicago, Drexel University, Fermilab, the 
Institute for Advanced Study, the Japan Participation Group, Johns 
Hopkins University, the Joint Institute for Nuclear Astrophysics, 
the Kavli Institute for Particle Astrophysics and Cosmology, the 
Korean Scientist Group, the Chinese Academy of Sciences (LAMOST), 
Los Alamos National Laboratory, the Max-Planck-Institute for Astronomy (MPIA),
the Max-Planck-Institute for Astrophysics (MPA), New Mexico State University, 
Ohio State University, University of Pittsburgh, University of Portsmouth, 
Princeton University, the United States Naval Observatory, and the 
University of Washington. We would also like to thank the referee for useful
coments which improved the paper. 
\end{acknowledgements}


\begin{thebibliography}{}


\bibitem[2009]{} Abazajian, K. N., Adelman-McCarthy, J. K., Ag\"ueros, M. A.,
et al. 2009, ApJS, 182, 543

\bibitem[2006b]{bell06b} Bellazzini, M., Ibata, R., Martin, N., Lewis, G. F.,
Conn, B., \& Irwin, M. J. 2006, MNRAS, 366, 865

\bibitem[2008]{} Bilir, S., Cabrera-Lavers, A., Karaali, S., Ak, S.,  Yaz, E.,
\& L{\'o}pez-Corredoira, M. 2008, PASA, 25, 69

\bibitem[2009]{}  Bournaud, F,  Elmegreen, B. G., \& Martig, M. 2009,  ApJ, 707, ,L1

\bibitem[2007]{butl07} Butler, D. J., Mart\'\i nez-Delgado, D., Rix, H.-W.,
Pe\~narrubia, J., \& de Jong, J. T. A., 2007, AJ, 133, 2274

\bibitem[2005]{} Conn, B. C., Lewis, G. F., Irwin, M. J., Ibata, R. A., 
Ferguson,  A. M. N., Tanvir, N., Irwin,  J. M., 2005, MNRAS, 362, 475 

\bibitem[2007]{conn07} Conn, B. C., Lane, R. R., Lewis, G. F., et al. 2007, 
MNRAS 376, 939

\bibitem[2008]{} Conn, B. C., Lane, R. R., Lewis, G. F., Irwin, M. J.,
Ibata, R. A., Martin, N. F., Bellazzini, M., \& Tuntsov, A. V. 2008,
MNRAS, 390, 1388 

\bibitem[2007]{jong07} de Jong, J. T. A., Butler, D. J., Rix, H.-W., Dolphin, A. E., 
Mart\'\i nez-Delgado, D. 2007, ApJ 662, 259

\bibitem[2008]{} Gilmore, G. \& Reid, N. 1983, MNRAS, 202, 1025
 

\bibitem[1995]{} Hammersley, P. L., Garz\'on, F., Mahoney, T., \& Calbet, X. 1995,
MNRAS, 273, 206
 
\bibitem[1994]{} Ibata, R. A., Gilmore, G., \& Irwin, M. J. 1994, Nature,
370, 194

\bibitem[2008]{} Ivezi\'c, Z., Sesar, B., Juri\'c, M., et al. 2008, 
ApJ, 684, 287

\bibitem[2007]{} Kalberla, P. M. W., Dedes, L., Kerp, J., \& Haud, U.
2007, A\&A, 469, 511

\bibitem[2006]{lope06} L\'opez-Corredoira, M., 2006, MNRAS 369, 1911

\bibitem[2002]{lope02} L\'opez-Corredoira, M., Cabrera-Lavers, A., Garz\'on, F.,
\& Hammersley, P. L., 2002, A\&A 394, 883

\bibitem[2007]{} L\'opez-Corredoira, M., Momany, Y., Zaggia, S., 
\& Cabrera-Lavers, A. 2007, A\&A 472, L47

\bibitem[2009]{} L\'opez-Corredoira, M. \& Betancort-Rijo, J. 2009,
A\&A, 493, L9

\bibitem[2004]{mart04} Martin, N.~F., Ibata, R.~A., Bellazzini, M., 
Irwin, M.~J., Lewis, G.~F., \& Dehnen, W. 2004, MNRAS, 348, 12 

\bibitem[2005]{} Mart\'\i nez-Delgado, D., Butler, D. J., Rix, H.-W.,
Franco, V. I., Pe\~narrubia, J., Alfaro, E. J., \& Dinescu, D. I. 2005,
ApJ, 633, 205

\bibitem[2004]{moma04} Momany, Y., Zaggia, S. R., Bonifacio, P., Piotto, G., De
Angeli, F., Bedin, L. R., \& Carraro, G. 2004, A\&A 421, L29

\bibitem[2006]{moma06} Momany, Y., Zaggia, S. R., Gilmore, G., Piotto, G.,
Carraro, G., Bedin, L. R., \& De Angeli, F. 2006, A\&A 451, 515

\bibitem[2002]{} Newberg, H. J., Yanny, B., Rockosi, C., et al. 2002,
ApJ, 569, 245

\bibitem[1992]{} Robin, A. C., Creze, M., \& Mohan, V. 1992, ApJ, 400, L25

\bibitem[2003]{} Robin, A. C., Reyl\'e, C., Derri\`ere, S., \& Picaud, S.
2003, A\&A, 409, 523

\bibitem[2003]{} Rocha-Pinto, H. J., Majewski, S. R., Skrutskie, M. F., 
\& Crane, J. D. 2003, ApJ, 594, L115

\bibitem[1998]{} Schlegel, D.J.,. Finkbeiner, D.P., \&  Davis, M. 1998, 
ApJ, 500, 525 

\end{thebibliography}
\end{document}